\documentclass[12pt]{article}
 \usepackage{graphicx}
 \usepackage[cp1251]{inputenc}
 \usepackage{rotating}
\begin{document}
 \noindent {\footnotesize\it Astronomy Letters, 2024, Vol. 50, No. 12}
 \newcommand{\dif}{\textrm{d}}

 \noindent
 \begin{tabular}{llllllllllllllllllllllllllllllllllllllllllllll}
 & & & & & & & & & & & & & & & & & & & & & & & & & & & & & & & & & & & & & &\\\hline\hline
 \end{tabular}

 \vskip 0.2cm
 \bigskip
 \centerline{\bf\Large  Is there an analogue of the Radcliffe wave between the}
 \centerline{\bf\Large  Carina-Sagittarius and the Scutum arms?}
 \bigskip
 \centerline{\bf   Vadim V. Bobylev\footnote[1]{bob-v-vzz@rambler.ru}   }
 \bigskip
 \centerline{\it Central (Pulkovo) Astronomical Observatory, Russian Academy of Sciences}
 \bigskip
 \bigskip
{The most complete sample of galactic maser sources and radio stars with trigonometric parallaxes, proper motions and radial velocities measured by the VLBI method has been compiled based on literature data. These sources are associated with young stars located in high mass star forming regions. The rotation parameters of the Galaxy have been determined based on 156 masers with relative parallax errors less than 10\%, located further than 3~kpc from the galactic center. The linear rotation velocity of the Galaxy at the solar distance $R_0$ is $V_0=243.9\pm3.9$~km s$^{-1}$. A very narrow chain of masers 3--4 kpc long, elongated in the $\sim40^\circ$ direction, passing from a segment of the Carina-Sagittarius spiral arm to the Scutum arm, has been studied. A number of authors have hypothesized that this is a possible analogue of the Radcliffe wave. In the present work, no noticeable periodic perturbations of vertical coordinates and velocities were found in this structure. On the other hand, on the diagram ``$\ln(R/R_0)-\theta$'' this chain of masers has the form of a segment of a logarithmic spiral with a pitch angle of $-48^\circ$. Perhaps this chain of masers belongs to a jet extending from the end of the bar, rotating rigidly with the angular velocity of the bar.
  }


 \bigskip
 \section*{INTRODUCTION}
Measurements of trigonometric parallaxes of stars are very important for the analysis of the structure and kinematics of the Galaxy. In the Gaia project (Prusti et al. 2016), trigonometric parallaxes, proper motions and radial velocities of millions of stars were determined (Gaia DR3, Valenari et al. 2022). However, at present, the accuracy of parallax determination in this catalog limits the study area to the radius of the solar region of about 3~kpc.

Sources of maser radiation and radio stars with trigonometric parallaxes and proper motions measured by the VLBI method currently play an important role in the study of the Galaxy. Young stars that have moved not far from their birthplace are of particular interest for determining the parameters of galactic rotation and spiral structure. VLBI measurements of about 270 such objects have been made to date with varying degrees of accuracy. Among them, there are about 160 masers with parallax measurement errors of about 10 microarcseconds on average and relative errors of such determinations of less than 10\%, which allows us to ``reach'' the center of the Galaxy.

We note a number of works that radio astronomers have devoted to VLBI measurements of masers and the study of the structure and kinematics of the Galaxy using the obtained material. These are, for example, large reviews by Reid et al. (2019), Hirota et al. (2020), and Immer and Rygl (2022).

The Radcliffe wave (Alves et al. 2020) has recently been discovered near the Sun. It is a narrow chain of molecular clouds, stretched into a line with a length of $\sim$2.7~kpc in the galactic $XY$ plane. Its main feature is a clearly visible wave-like nature of the cloud distribution in the vertical direction. The maximum value of the $Z$ coordinate is $\sim160$~pc, observed in the immediate vicinity of the Sun.

A fairly large number of publications have already been devoted to the study of the Radcliffe wave.
The wave-like behavior of vertical coordinates has been confirmed in the distribution of interstellar dust~(Lallement et al.~2022; Edenhofer et al.~2024), molecular clouds~(Zucker et al.~2023), masers and radio stars~(Bobylev et al.~2022),
T-Tauri stars~(Li, Chen~2022), massive OB stars~(Donada, Figueras~2021; Thulasidharan et al. 2022), as well as young open star clusters~(Donada, Figueras 2021). The Radcliffe wave also exhibits periodicity in vertical velocities with an amplitude of 5--10~km s$^{-1}$ (Bobylev et al.~2022; Konietzka et al.~2024). Thus, the Radcliffe wave is a reflection of a real physical process that led to a perturbation of the vertical coordinates and vertical velocities of a stellar-gas structure with a mass of about $3\times10^6~M_\odot$.

In the paper by Mai et al. (2023), devoted to VLBI measurements of trigonometric parallaxes of masers, an interesting hypothesis was put forward about the possibility of the existence of an analogue of the Radcliffe wave elsewhere in the Galactic disk. Namely, an unusually long, very narrow chain of sources is noticeable in the distribution of masers in projection on the galactic plane $XY$, extended in the direction $l\sim40^\circ$, passing from a segment of the Carina-Sagittarius spiral arm to the Scutum arm.

The aim of this work is to compile the most complete list of data on masers from literary sources with VLBI measurements of their trigonometric parallaxes, proper motions and radial velocities. To determine a number of kinematic parameters of the Galaxy's rotation from these data. The aim is also to test the idea of Mai et al. (2023). To do this, it is necessary to determine the presence of periodic perturbations of vertical coordinates and velocities in the formation of masers indicated by them.

\begin{figure}[t]
{ \begin{center}
 \includegraphics[width=0.45\textwidth]{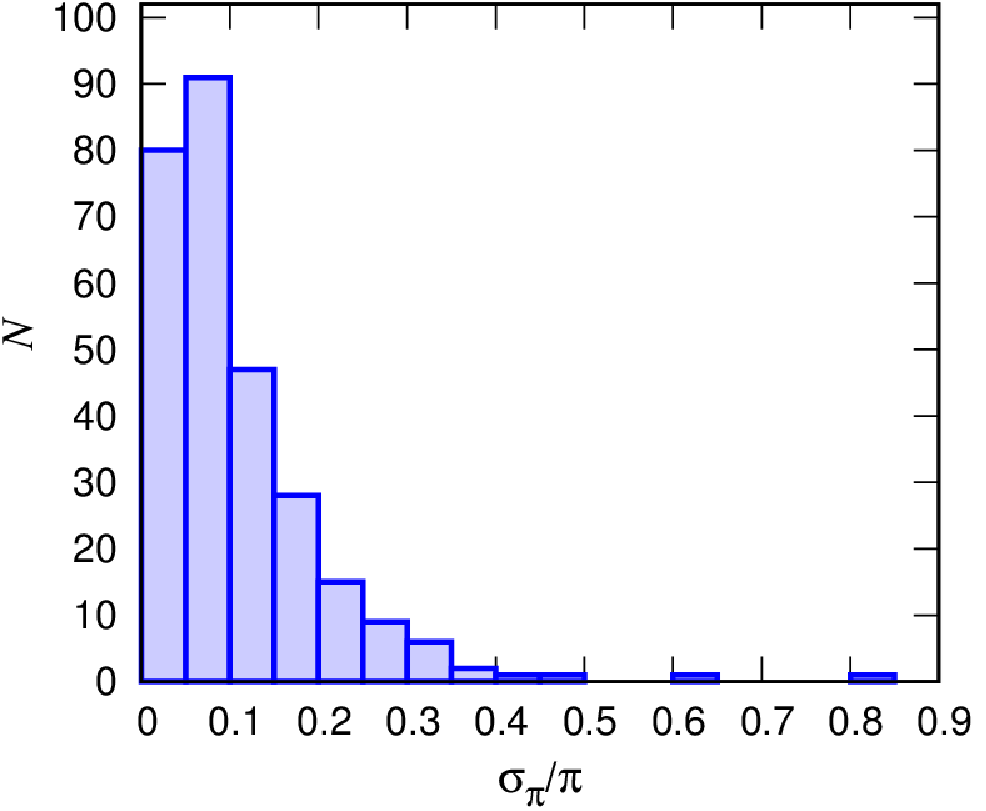}
\caption{The number of masers and radio stars depending on the relative error of VLBI measurements of their trigonometric parallaxes.}
 \label{f-1}
\end{center}}
\end{figure}

\section*{DATA}
Maser emission occurs in the immediate vicinity of young, forming stars, as well as already evolved stars, pumped either by strong infrared radiation or by strong gas collisions in disks, flows in jets or winds. A special feature of masers is that almost all their energy is emitted in a few molecular lines. These include, for example, hydroxyl (OH) masers with a frequency of 1.6 GHz, methanol (CH$_3$OH) masers with a frequency of 6.7 GHz and 12.2 GHz, water vapor (H$_2$O) masers with a frequency of 22 GHz, or silicon monoxide (SiO) masers with a frequency of 43 GHz.

The use of the VLBI method to measure the trigonometric parallaxes of galactic masers with relative errors on average less than 10\% has made them first-class objects for studying the structure and kinematics of the Galaxy. The value of radio observations is that they are not hampered by the absorption of radiation by interstellar dust. Of greatest interest are maser sources associated with young stars and protostars that are located in high mass star forming regions. A sample of 199 such sources is presented in the paper by Reid et al. (2019). VLBI observations were carried out within the framework of the BeSSeL (The Bar and Spiral Structure Legacy Survey \footnote {http://bessel.vlbi-astrometry.org}) project. The most important contributor here is the American VLBA array, consisting of ten 25-meter antennas with a maximum baseline of more than 8000~km. The observations cover frequencies of 6.7~GHz and 12.2~GHz with methanol maser transitions, as well as water vapor maser transitions with a frequency of 22.2~GHz.

Another contributor to the BeSSeL program is the European VLBI Network (EVN). Here the longest baselines are about 9000 km, and the largest in the array is the 100-meter antenna in Effelsberg. Observations are carried out at frequencies from 6.7 GHz to 22.2 GHz.

 {\begin{table}[t]                                    
 \caption[]
 {\small\baselineskip=1.0ex
Number of major and radio stars with relative parallax errors}
 \label{t-1}
 \begin{center}\begin{tabular}{|c|c|c|}\hline
 $\sigma_\pi/\pi$  &  R$>$0~kpc& 3$<$R$<$14~kpc\\\hline
 30\% &   267 &   259 \\
 20\% &   243 &   236 \\
 15\% &   215 &   210 \\
 10\% &   168 &   164 \\\hline
 \end{tabular}\end{center}\end{table}}

Finally, VLBI observations of masers are carried out in Japan under the VERA (VLBI Exploration of Radio Astrometry \footnote[2]{http://veraserver.mtk.nao.ac.jp}) program, which is also part of the BeSSeL program. The interferometer consists of four 20-meter antennas located throughout Japan, which provides a baseline length of 1020 to 2270 km. Observations of H$_2$O masers are carried out at a frequency of 22.2 GHz, and less frequently, SiO masers at a frequency of 43.1 and 42.8 GHz. A unique property of the VERA antennas is the dual-beam receiving system, which allows simultaneous tracking of a pair of maser targets and phase reference sources. In all other programs (VLBA, EVN, etc.), observations of reference extragalactic objects are made at the beginning and end of the session by re-aiming the antennas, which then requires additional efforts to account for atmospheric distortions. Note that the higher the observation frequency, the better the astrometric accuracy. Thus, VLBI observations performed by the VERA program are the most accurate compared to observations obtained within the framework of other programs. Hirota et al. (2020) described a catalog of 99 sources for which trigonometric parallaxes and proper motion components were obtained exclusively by the VERA program.

Recently, the East Asian VLBI Network (EAVN) \footnote[3]{https://radio.kasi.re.kr/eavn/main$\_$eavn.php }, a joint network of the Korean VLBI Network (KVN), Chinese VLBI Network (CVN), and Japanese VERA, has been launched. Currently, EAVN consists of 21 telescopes that conduct observations of H$_2$O masers at a frequency of 22.2~GHz (Akiyama et al. 2022). The first results of determining the trigonometric parallax of the maser source are presented in the paper by Sakai et al. (2023).

Results of VLBI measurements of masers have also appeared, obtained using the Long Baseline Array (LBA) radio interferometer in Australia (Krishnan  et al. 2015). The interferometer consists of large-diameter antennas (more than 20 meters), with the help of which methanol masers are observed at a frequency of 6.7 GHz.

For each maser in the lists of Reid et al. (2019) and Hirota et al. (2020), the values of the measured equatorial coordinates, trigonometric parallax, and two components of proper motion are given. The radial velocities taken from various literary sources are also given, and an extensive bibliography is provided. The lists of Reid et al. (2019) and Hirota et al. (2020) overlap, so in this paper we have compiled a common list of data without overlaps.

To this general list we have added new measurements performed after 2020 (Sakai et al. 2020; 2023; Ortiz-Le\'on et al. 2020; 2023; Xu et al. 2021; Bian et al. 2022; Mai et al. 2023). The most interesting are the measurements of four maser sources performed from the southern hemisphere (Hyland et al. 2023; 2024) using the Australian radio interferometer. These objects are located in the fourth galactic quadrant, where previously there were virtually no measurement data.

In addition to maser sources, our list includes radio stars whose VLBI observations were performed not in molecular lines (as in the case of masers), but in the continuum (Ortiz-Le\'on et al. 2018; Galli et al. 2018). Most of these are low-mass T Tauri stars located in the Gould Belt region. Currently, for approximately 60 such stars observed at frequencies of 5~GHz and 8~GHz by the GOBELINS program (Gould's Belt Distances Survey, Ortiz-Le\'on et al. 2017), their absolute trigonometric parallaxes and proper motions have been measured, and their radial velocities are known.

As the measurement material accumulates, works devoted to the analysis of the structure and kinematics of the Galaxy using data on masers in high mass star forming regions periodically appear. We can note the works of Reid et al. (2009), Bobylev, Bajkova (2010), Honma et al. (2012), Rastorguev et al. (2017), Reid et al. (2019), Hirota et al. (2020), Bobylev, Bajkova (2022).

\begin{figure}[t]
{ \begin{center}
 \includegraphics[width=0.95\textwidth]{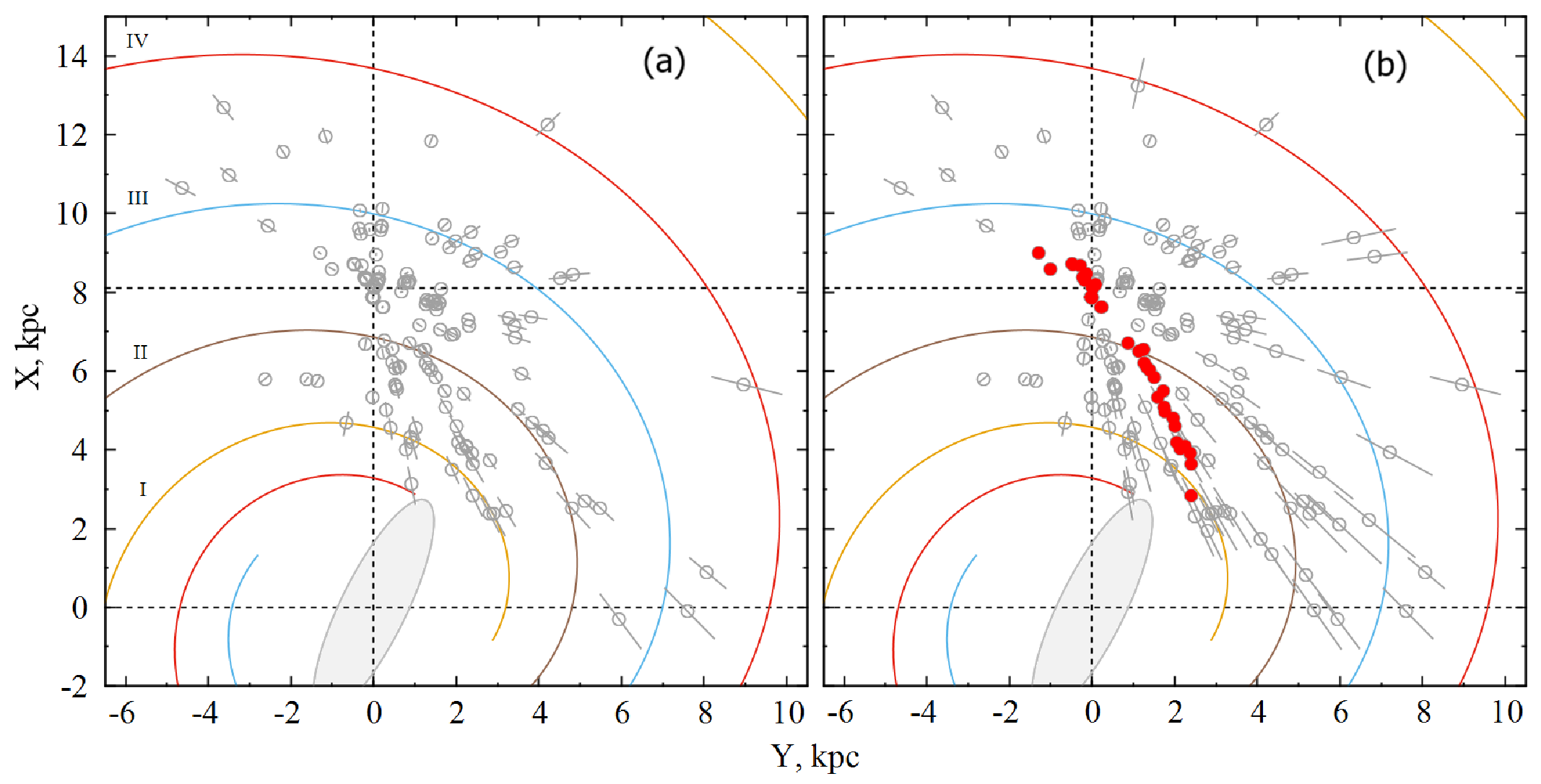}
\caption{Distribution of 164 masers and radio stars with parallax measurement errors less than $10\%$ in projection onto the galactic plane $XY$~(a), distribution of 210 masers with parallax measurement errors less than $15\%$~(b).}
 \label{f-2-XY}
\end{center}}
\end{figure}

At present, the list contains only 308 results of determination of VLBI parallaxes and proper motions of masers and radio stars (associated with protostars and young stars located in high mass star forming regions).

Figure~\ref{f-1} shows a histogram of the distribution of stars depending on the relative error in determining the parallax for the entire sample of objects under study, and Table~\ref{t-1} shows the number of objects depending on the relative error in determining the parallax $\sigma_\pi/\pi$ and their position in the Galaxy (depending on the distance $R$). The statistics are given both for the entire sample and for the distance interval $R:[3-14]$~kpc, which is of great interest for studying the rotation of the Galaxy (e.g., Reid et al. 2009; 2019), as well as the parameters of the spiral structure (e.g., Rastorguev et al. 2017; Reid et al. 2019; Bobylev et al. 2020; Bobylev, Bajkova 2022). The most important condition here is $R<3$~kpc, which greatly limits the influence of the galactic bar on the motion of stars.

Fig.~\ref{f-2-XY}(a) shows the distribution of 164 masers and radio stars with parallax measurement errors less than $10\%$ in projection onto the galactic plane $XY$. Fig.~\ref{f-2-XY}(b) shows the distribution of 210 masers with parallax measurement errors less than $15\%$ in projection onto the galactic plane $XY$, where the chain of masers that we want to check for perturbations of the vertical coordinates $Z$ and velocities $W$ is highlighted by red circles.
This figure uses a coordinate system in which the $X$ axis is directed from the center of the Galaxy to the Sun, and the direction of the $Y$ axis coincides with the direction of rotation of the Galaxy. The four-arm spiral pattern with a pitch angle of $i=-13^\circ$ is given according to the work of Bobylev, Bajkova (2014), here it is constructed with a value of $R_0=8.1$~kpc, the following four spiral arms are numbered with Roman numerals: I~--- Scutum, II~--- Carina-Sagittarius, III~--- Perseus and IV~--- Outer arm.

 \section*{METHODS}\label{method}
From observations we have three stellar velocity components: the line-of-sight velocity $V_r$ and the two tangential velocity components $V_l=4.74r\mu_l\cos b$ and $V_b=4.74r\mu_b$ along the Galactic longitude $l$ and latitude $b,$ respectively, expressed in km s$^{-1}$. Here, the coefficient 4.74 is the ratio of the number of kilometers in an astronomical unit to the number of seconds in a tropical year and $r$ is the stellar heliocentric distance in kpc, $r=1/\pi$. The proper motion components $\mu_l\cos b$ and $\mu_b$ are expressed in mas yr$^{-1}$. The velocities $U,V,W$ directed along the
rectangular Galactic coordinate axes are calculated via the components $V_r, V_l, V_b$:
 \begin{equation}
 \begin{array}{lll}
 U=V_r\cos l\cos b-V_l\sin l-V_b\cos l\sin b,\\
 V=V_r\sin l\cos b+V_l\cos l-V_b\sin l\sin b,\\
 W=V_r\sin b                +V_b\cos b,
 \label{UVW}
 \end{array}
 \end{equation}
where the velocity $U$ is directed from the Sun toward the Galactic center, $V$ is in the direction of Galactic
rotation, and $W$ is directed to the north Galactic pole. We can find two velocities, $V_R$ directed radially
away from the Galactic center and the velocity $V_{\rm circ}$ orthogonal to it pointing in the direction of Galactic
rotation from the following expressions:
 \begin{equation}
 \begin{array}{lll}
  V_{circ}= U\sin \theta+(V_0+V)\cos \theta, \\
       V_R=-U\cos \theta+(V_0+V)\sin \theta,
 \label{VRVT}
 \end{array}
 \end{equation}
where the position angle $\theta$ obeys the relation $\tan\theta=y/(R_0-x)$, $x,y,z$ are the rectangular heliocentric coordinates of the star (the velocities $U,V,W$ are directed along the corresponding $x,y,z$ axes), and $V_0$ is the linear rotation velocity of the Galaxy at the solar distance $R_0$.

 \begin{table}[t] \caption[]{\small
Kinematic parameters found from 164 maser sources located in the $R>3$~kpc region of the Galaxy. }
  \begin{center}  \label{t:164}
  \small\begin{tabular}{|l|r|r|r|r|r|}\hline
                        Parameters  & $V_r+V_l+V_b$  &      $V_l+V_b$ \\\hline
   $U_\odot,$  km s$^{-1}$   & $ 9.42\pm0.81$ & $ 9.61\pm1.32$ \\
   $V_\odot,$  km s$^{-1}$   & $13.17\pm0.84$ & $11.49\pm0.96$ \\
   $W_\odot,$  km s$^{-1}$  & $ 8.96\pm0.73$ & $ 8.49\pm0.62$ \\
        $\Omega_0,$  km s$^{-1}$ kpc$^{-1}$   & $ 30.11\pm0.31$ & $ 29.66\pm0.36$ \\
  $\Omega^{'}_0,$  km s$^{-1}$ kpc$^{-2}$  & $-4.333\pm0.067$& $-3.917\pm0.096$\\
 $\Omega^{''}_0,$  km s$^{-1}$ kpc$^{-3}$  & $ 0.837\pm0.034$& $ 0.648\pm0.043$\\
      $\sigma_0,$ km s$^{-1}$          &          9.0    &             7.9 \\
           $V_0,$ km s$^{-1}$             &   $243.9\pm3.9$ &   $240.2\pm4.1$ \\\hline
\end{tabular}\end{center} \end{table}

\subsection*{Determination of the rotation parameters of the Galaxy}
To determine the parameters of the galactic rotation curve, we use the equations obtained from the Bottlinger formulas, in which the angular velocity $\Omega$ is expanded into a series up to terms of the second order of smallness $r/R_0:$
\begin{equation} \begin{array}{lll}
 V_r=-U_\odot\cos b\cos l-V_\odot\cos b\sin l\\
 -W_\odot\sin b+R_0(R-R_0)\sin l\cos b\Omega^\prime_0\\
 +0.5R_0(R-R_0)^2\sin l\cos b\Omega^{\prime\prime}_0,
 \label{EQ-1}
 \end{array} \end{equation}
\begin{equation} \begin{array}{lll}
 V_l= U_\odot\sin l-V_\odot\cos l-r\Omega_0\cos b\\
 +(R-R_0)(R_0\cos l-r\cos b)\Omega^\prime_0\\
 +0.5(R-R_0)^2(R_0\cos l-r\cos b)\Omega^{\prime\prime}_0,
 \label{EQ-2}
 \end{array}\end{equation}
\begin{equation} \begin{array}{lll}
 V_b=U_\odot\cos l\sin b + V_\odot\sin l \sin b\\
 -W_\odot\cos b-R_0(R-R_0)\sin l\sin b\Omega^\prime_0\\
    -0.5R_0(R-R_0)^2\sin l\sin b\Omega^{\prime\prime}_0,
 \label{EQ-3}
 \end{array} \end{equation}
where $R$ is the distance from the star to the rotation axis of the Galaxy $R^2=r^2\cos^2 b-2R_0 r\cos b\cos l+R^2_0.$ The velocities $(U,V,W)_\odot$ are the average group velocity of the sample, they reflect the peculiar motion of the Sun, therefore they are taken with opposite signs. $\Omega_0$ is the angular velocity of rotation of the Galaxy at the solar distance $R_0$, the parameters $\Omega^{\prime}_0$ and $\Omega^{\prime\prime}_0$~--- the corresponding derivatives of the angular velocity of rotation, $V_0=R_0\Omega_0$.

We take $R_0$ to be $8.1\pm0.1$~kpc. This value was derived as a weighted average from a large number of modern individual estimates in Bobylev, Bajkova (2021).

The solution of conditional equations of the form~(\ref{EQ-1})--(\ref{EQ-3}) is sought by the least squares method (LSM). As a result, we obtain an estimate of the following six unknowns: $(U,V,W)_\odot,$ $\Omega_0$, $\Omega^{\prime}_0$ and $\Omega^{\prime\prime}_0$. Note that the velocities $U,$ $V$ and $W$ in equations~(\ref{VRVT}) are freed from the peculiar velocity of the Sun $U_\odot,$ $V_\odot$ and $W_\odot$ with the values found as a result of the LSM solution of kinematic equations of the form~(\ref{EQ-1})--(\ref{EQ-3}).

  \subsection*{Spiral density wave}
The position of a star in a logarithmic spiral wave is generally written as follows:
 \begin{equation}
 R=a_0 e^{(\theta-\theta_0)\tan i},
 \label{spiral-1}
 \end{equation}
where $R$~--- the distance from the center of the Galaxy to the star, $\theta$~--- the position angle of the star, $\theta_0$~--- an arbitrarily chosen initial angle, which we set equal to zero; $a_0$~--- the place where the spiral intersects the $X$ axis; $i$~--- the pitch angle of the spiral pattern ($i<0$ for a twisting spiral), which is related to the other parameters as follows:
 \begin{equation}
\tan (|i|)=m\lambda/(2\pi R_0),
 \label{sp-lambda}
 \end{equation}
where $m$~--- the number of spiral arms, $\lambda$~--- the wavelength equal to the distance (in the radial direction) between the segments of the spiral arms in the circumsolar region. After dividing the left and right sides of the equation~(\ref{spiral-1}) by $R_0$ and taking the logarithm, we obtain the relation
 \begin{equation}
 \ln (R/R_0)=\ln (a_0/R_0)+ \theta\tan i,
 \label{spiral-2}
 \end{equation}
which is the equation of a straight line. Then from the diagram ``$\ln (R/R_0)-\theta$'' we can estimate the value of $a_0$ and $i$.

\begin{figure}[t]
{ \begin{center}
 \includegraphics[width=0.9\textwidth]{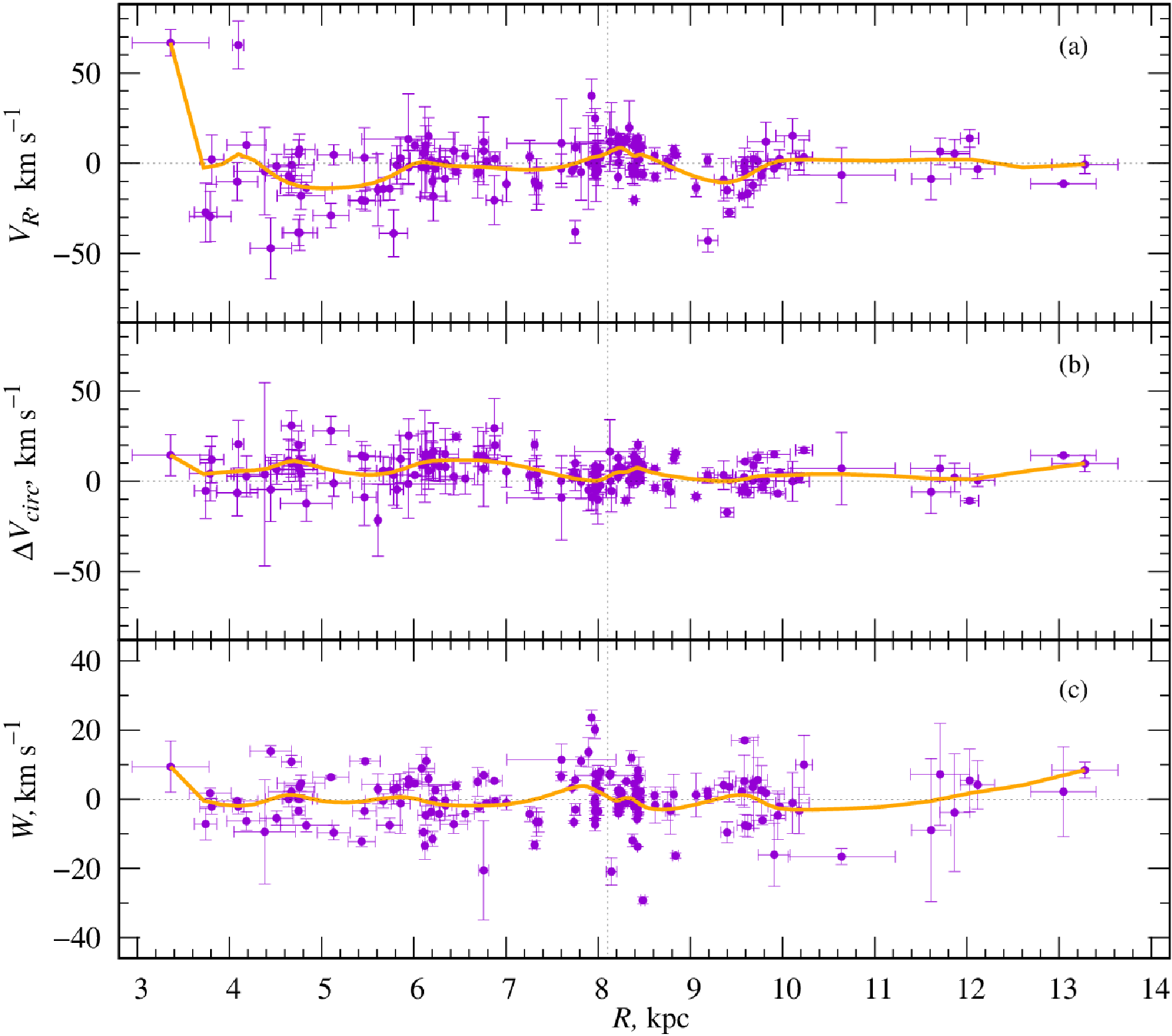}
\caption{Velocities of masers and radio stars with relative errors in determining parallaxes less than 10\% depending on the distance $R$: a) radial; b) residual tangential; and c) vertical.}
 \label{f-3}
\end{center}}
\end{figure}
\begin{figure}[t]
{ \begin{center}
  \includegraphics[width=0.9\textwidth]{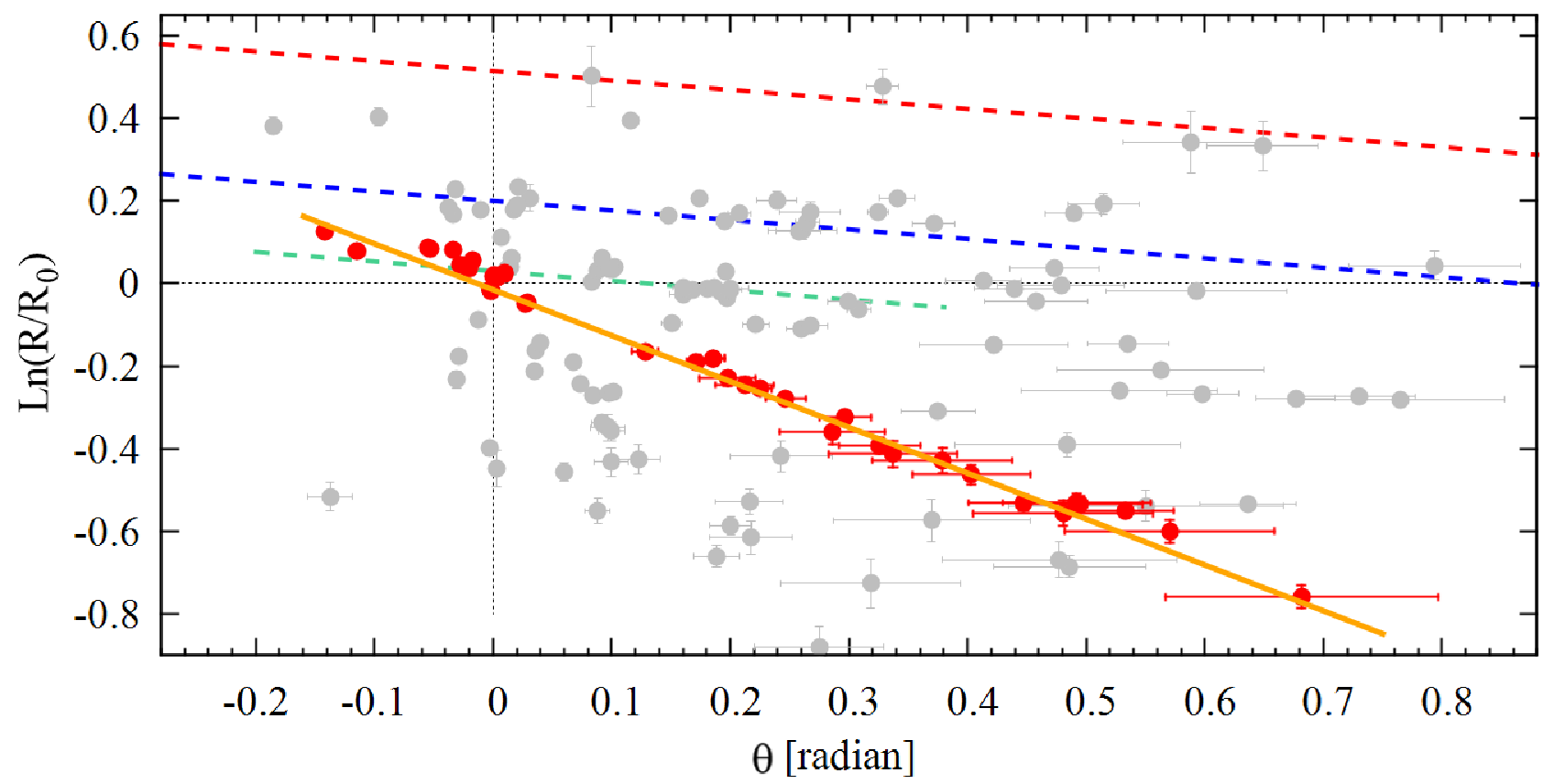}
\caption{Galactocentric distances of masers of the form $\ln (R/R_0)$ depending on the position angles $\theta$, the dotted lines are drawn at an angle to the horizontal axis $-13^\circ$ for the Outer Arm (red line), Perseus (blue line), Orion (green line), and for the chain of red dots at an angle $-48^\circ$ (orange solid line).}
 \label{f-4}
\end{center}}
\end{figure}
\begin{figure}[t]
{ \begin{center}
  \includegraphics[width=0.8\textwidth]{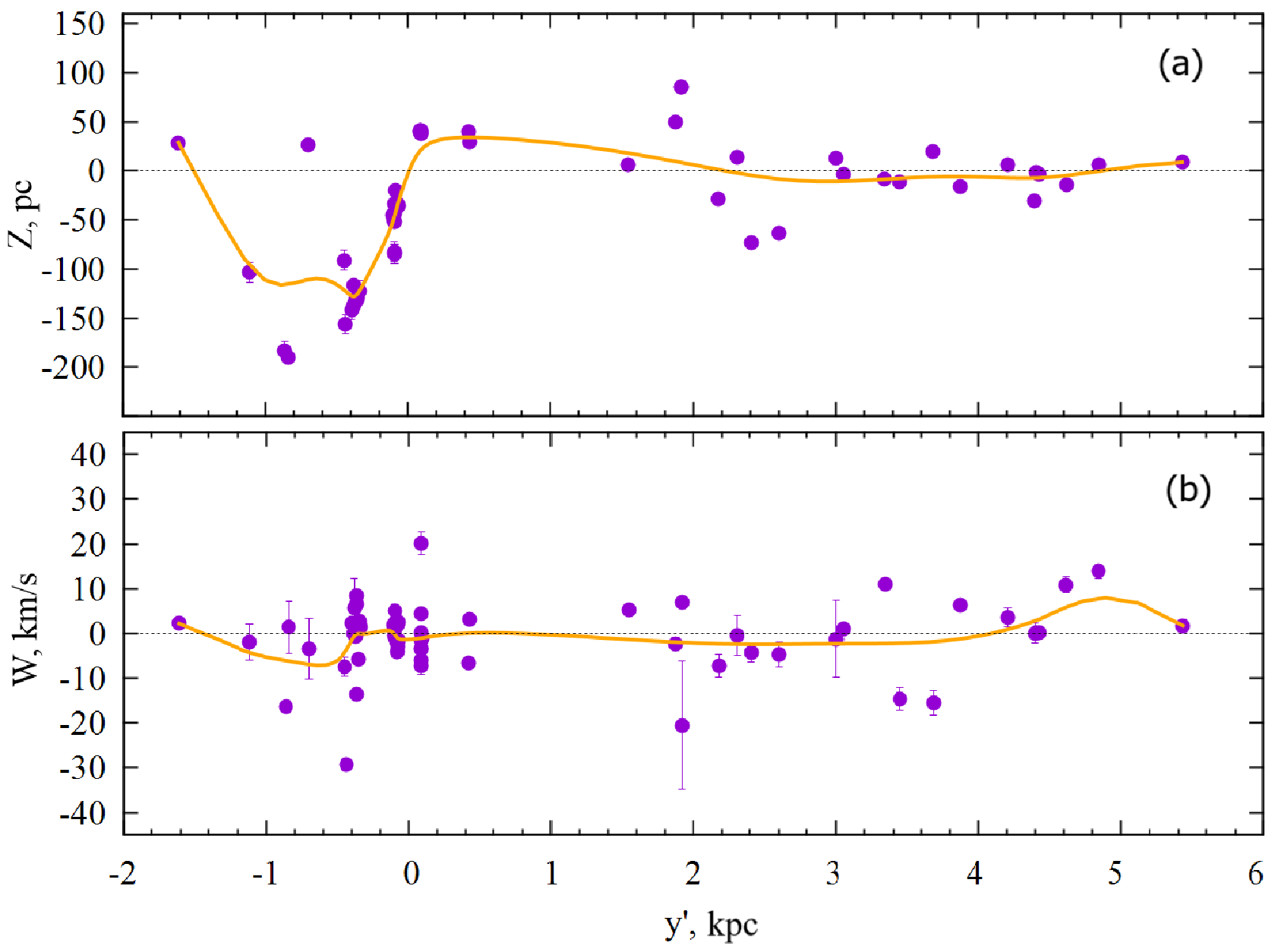}
\caption{Vertical coordinates~(a) and velocities~(b) of the maser chain along the $y'$ axis.}
 \label{f-5}
\end{center}}
\end{figure}
\begin{figure}[t]
{ \begin{center}
  \includegraphics[width=0.6\textwidth]{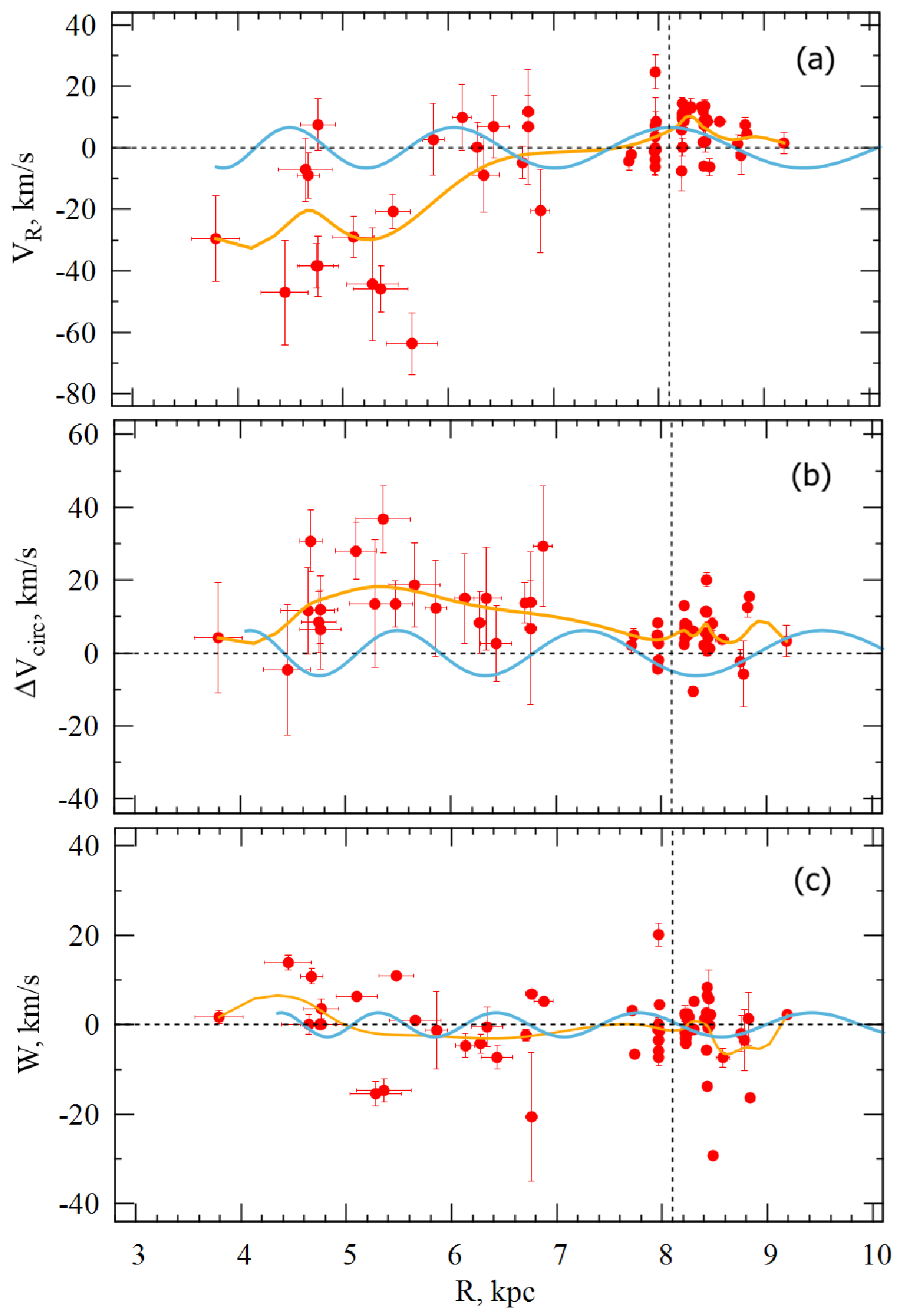}
\caption{Velocities of the selected chain of masers and radio stars depending on the distance $R$: a)~radial; b)~residual tangential and c)~vertical, see text.}
 \label{f-6}
\end{center}}
\end{figure}

 \section*{RESULTS AND DISCUSSION}
  \subsection*{Rotation parameters of the Galaxy}
  \label{rotation}
Two methods were used to find the least squares solution of the kinematic equations. In the first case, three velocities were used -- $V_r$, $V_l$ and $V_b$ (option $V_r+V_l+V_b$), and in the second case, only the proper motions of the masers were used (option $(V_l+V_b$)). The results are shown in the table~\ref{t:164}.

In the work of Bobylev and Bajkova (2022) for 150 masers under similar conditions for the $V_r+V_l+V_b$ variant the following estimates were obtained:
$(U,V,W)_\odot=(9.15,12.81,8.93)\pm(0.86,0.86,0.75)$~km s$^{-1}$,
$\Omega_0=30.18\pm0.38$~km s$^{-1}$ kpc$^{-1}$ ,
$\Omega^{'}_0=-4.368\pm0.077$~km s$^{-1}$ kpc$^{-2}$   and
$\Omega^{''}_0=0.845\pm0.037$~km s$^{-1}$ kpc$^{-3}$ ,
where the unit weight error $\sigma_0$ was 8.9~km s$^{-1}$ and $V_0=244.4\pm4.3$~km s$^{-1}$ (for the adopted $R_0=8.1\pm0.1$~kpc). Comparison of the values of these parameters with the corresponding ones from Table~\ref{t:164} shows that in the present work all estimates were obtained with smaller errors. The improvement of the solution occurred due to the use of a larger number of masers in the solutions.

The radial $V_R$, residual tangential $\Delta V_{circ}$ and vertical $W$ velocities of masers and radio stars as functions of distance $R$ are shown in Fig.~\ref{f-3}. The averaged velocity values are shown in the orange lines in the figures. All three panels of the figure clearly show periodicities reflecting velocity perturbations caused by the galactic spiral density wave. The parameters of such perturbations (amplitudes and wavelengths) were determined in the work of Bobylev and Bajkova (2022) using spectral analysis of three types of velocities --- $V_R$, $\Delta V_{circ}$ and $W$. Depending on the type of velocities, the values of the disturbance parameters were found to be as follows: $\lambda_{R,\theta,W}=(2.1,2.1,2.6)$~kpc and the disturbance amplitudes $f_{R,\theta,W}=(6.7,2.6,5.2)$~km s$^{-1}$. In the work of Bobylev and Bajkova (2022), very good agreement was shown between the curves obtained by means of a moving average (orange lines in Fig.~\ref{f-3}) and the periodic curves found as a result of spectral analysis.

  \subsection*{Diagram ``$\ln (R/R_0)-\theta$''}
To construct the ``$\ln (R/R_0)-\theta$'' diagram, which is given in Fig.~\ref{f-4}, a sample of masers with parallax measurement errors less than 15\% was used. For segments of the Orion, Perseus, and Outer spiral arms, lines with a pitch angle of $-13^\circ$ were drawn. For the chain of masers marked with red dots in the figure, it was found that they are located at an angle of $-48^\circ$. For them, the estimate $\ln (a/R_0)=-0.015$ was also obtained. In what follows, to select this chain of masers, we use a selection zone with a width of $\ln (R/R_0)=0.12$ (0.97~kpc). For this, we use the restriction on the angle $-0.2<\theta<0.8$~radians and two restrictive lines on the diagram:
 \begin{equation}
 \begin{array}{lll}
 \ln (R/R_0)= ~0.045+ \theta\tan (-48^\circ),\\
 \ln (R/R_0)= -0.075+ \theta\tan (-48^\circ).
 \label{spiral-3}
 \end{array}
 \end{equation}
The masers selected in this way, of which there were 65 sources, are marked with red circles in Fig.~\ref{f-2-XY}(b). At first we wanted to look only at the properties of a shorter chain of stars, i.e., those in the interval $0.1<\theta<0.8$~radians, but it turned out that including stars from the Sun's immediate environment is more informative.

  \subsection*{Radcliffe Wave and the Carina-Scutum Chain}
To determine the presence of periodic perturbations of the vertical coordinates and velocities of the selected masers, a transition was made (in the heliocentric coordinate system) to the primed $y'$ axis at an angle of $48^\circ$ as follows
\begin{equation}
 y'= y\cos{48^\circ}+x\sin{48^\circ}.
 \label{y'-48}
\end{equation}
This approach is usually used in the study of the Radcliffe wave (Alves et al. 2020; Bobylev et al. 2022; Konietzka et al. 2024). The dependences of the vertical coordinates and velocities of the selected 65 masers on the $y'$ coordinates are given in Fig. \ref{f-5}.

From Fig.~\ref{f-5} it is evident that masers located in the interval $1<y'<6$~kpc do not exhibit noticeable periodic perturbations of vertical coordinates and velocities characteristic of the Radcliffe wave.

In the region $-2<y'<1$~kpc in this figure there are masers under the influence of the Radcliffe wave. The manifestation of this wave is clearly visible in the $Z$ coordinates. The influence of the Radcliffe wave is reduced here (we see a different value of the wavelength here). It is well known that the Radcliffe wave is oriented at an angle of $25^\circ-30^\circ$ to the $y$ axis. But the disturbance amplitude $Z\sim150$~pc (Fig.~\ref{f-5}(a)) is in complete agreement with the known Radcliffe wave.

On the other hand, the narrow chain of selected masers (Fig.~\ref{f-4}) looks like a segment of a logarithmic spiral with a pitch angle of $-48^\circ$. If this is real, then we can assume that such a large-scale spiral wave could provoke the formation of the Gould Belt, and could also excite a wave in vertical coordinates and velocities in the Radcliffe wave. This hypothesis is all the more interesting because numerical modeling of the spiral structure of the Galaxy in the presence of a bar provides arguments in its support. For example, in Fig.~3 from the work of Li et al. (2022), constructed based on the results of such modeling, we can see a jet extending from the ends of the bar, similar to our case. In the model of Li et al. (2022), the value of the angular velocity of the bar was taken to be $\Omega_b=37.5$~km s$^{-1}$ kpc$^{-1}$. Let us assume that the maser chain we have discovered is a similar jet that rotates rigidly with the angular velocity of the bar. This means that in the solar region the jet runs into the Radcliffe wave matter, since in this case the Radcliffe wave is located further from the center of the Galaxy than the corotation radius of the bar. This agrees with both the radial and tangential motion of the Radcliffe wave, recently discovered in the work of Konietzka et al. (2024). Another argument in support of this hypothesis is that the described maser chain originates from the end of a long bar. According to definitions by various authors (e.g., Benjamin et al. 2005; Cabrera-Lavers et al. 2008; Wegg et al. 2015), the long bar is oriented at an angle of $30^\circ$ to $45^\circ$ to the $X$ axis, and its half-length is between 4 and 5~kpc.

Fig.~\ref{f-6} shows the radial, residual tangential and vertical velocities of the selected maser chain as functions of the distance $R$. In this figure, the blue line shows the periodic curves found in the work of Bobylev, Bajkova (2022) for a sample of masers based on spectral analysis. These curves, the parameters of which have already been listed when discussing the parameters of the rotation of the Galaxy, reflect the influence of the galactic spiral density wave. The orange line shows the average values of the velocities. As can be seen from the figure, in the region of the Scutum arm ($R\sim5$~kpc), the velocities $V_R$ and $\Delta V_{circ}$ have significant deviations from the blue curve. This indicates an unusual kinematics of the entire group of sources in this region.

The value of the pitch angle of the spiral close to $-48^\circ$ is very rare, but it is found in other galaxies (see, for example, Fig.~9 in Yu, Ho 2019). The spiral pattern with the pitch angle $i=-48^\circ$ for small values of $m$ should have a large wavelength $\lambda$. Using the relation~(\ref{sp-lambda}) we can estimate the value of $\lambda$ for $i=-48^\circ$ and $R_0=8.1$~kpc:
$\lambda=28$~kpc for $m=2$, or $\lambda=14$~kpc for $m=4$, etc.

Models of the spiral structure of the Galaxy consisting of a superposition of several spiral patterns have been discussed many times in the literature. For example, the model of L\'epine et al. (2001) is known about the coexistence of a two-arm and four-arm spiral patterns (with pitch angles of $6^\circ$ and $12^\circ$, respectively) at the solar radius. Englmaier et al. (2008) proposed a model in which a two-arm pattern is realized inside the solar circle, and in the outer part of the Galaxy the spiral pattern already becomes four-arm (see also the work of these authors Pohl et al.~2008).

It is also interesting to note the works of Griv et al. (2017; 2021), where, based on the kinematic analysis of samples of various young objects, a conclusion was made about the implementation of a single-arm mode ($m=1$) in the Galaxy. That is, in such a model $\lambda=2\pi R_0\tan |i|$. Then, applied to our case (Bobylev, Bajkova 2014), we find $\lambda\sim 12$~kpc for $m=2$, $i=-13^\circ$ and $R_0=8.1$~kpc, or $\lambda \sim24$~kpc for $m=4$, $i=-13^\circ$ and $R_0=8.1$~kpc. However, Griv et al. (2017; 2021) obtained quite ordinary values of the density wave parameters in the solar neighborhood at $m=1$: $i\approx-2^\circ$ and $\lambda\approx2$~kpc.

\section*{CONCLUSIONS}
Based on literature data, the most complete sample of galactic maser sources and radio stars with trigonometric parallaxes, proper motions and radial velocities measured by the VLBI method has been compiled to date. These masers and radio stars are very young objects associated with protostars, stars that have not reached the main sequence, and in some cases with very massive young stars. All of them are closely associated with high mass star forming regions.

As a result of the joint solution of the system of kinematic equations using 164 masers with errors in measuring their trigonometric parallaxes less than 10\%, located in the region of the Galaxy $R>3$~kpc, the components of the group velocity $(U,V,W)_\odot=(9.15,12.81,8.93)\pm(0.86,0.86,0.75)$~km s$^{-1}$ and the following parameters of the angular velocity of rotation of the Galaxy were found:
$\Omega_0=30.11\pm0.31$~km s$^{-1}$ kpc$^{-1}$,
$\Omega^{'}_0=-4.333\pm0.067$~km s$^{-1}$ kpc$^{-2}$ and
$\Omega^{''}_0=0.837\pm0.034$~km s$^{-1}$ kpc$^{-3}$,
where the linear velocity of rotation of the Galaxy at the solar distance was $V_0=243.9\pm3.9$~km s$^{-1}$ for the accepted value $R_0=8.1\pm0.1$~kpc.

A very narrow chain of masers 3--4 kpc long, elongated in the $l\sim40^\circ$ direction, passing from a segment of the Carina-Sagittarius spiral arm to the Scutum arm, was studied. The hypothesis of Mai et al. (2023) that this chain of masers may be an analogue of the Radcliffe wave was tested. In the present work, no significant periodic perturbations of vertical coordinates and velocities in this structure, characteristic of the Radcliffe wave, were found. That is, the chain of masers between segments of the Carina and Scutum spiral arms is not an analogue of the Radcliffe wave.

In the course of the analysis it was found that the narrow chain of masers, which already stretches from the Sun (covers the Gould Belt region) to the section of the Scutum arm, looks like a section of a logarithmic spiral with a pitch angle of $-48^\circ$. It is possible that such a large-scale spiral density wave could have provoked the formation of the Gould Belt, as well as excited a wave in vertical coordinates and velocities in the Radcliffe wave. It would be interesting to test this hypothesis on a larger sample. At present, there is a lack of highly accurate data in terms of parallax measurements to test it. For example, the distances to numerous clouds of neutral hydrogen H~I, distributed practically over the entire disk of the Galaxy, have not been determined using a very accurate kinematic method. Distances to classical Cepheids are measured with an accuracy of 5--10\%, but they are distributed with a large dispersion relative to the centers of the spiral arms, so one cannot expect a good concentration of them to the narrow chain described in this paper. Hope for the final version of the Gaia catalog.

Our working hypothesis is that the detected maser chain is a jet-like structure emanating from the end of the bar, which is rigidly rotating at the angular velocity of the bar.

\bigskip\medskip{REFERENCES}\medskip {\small

1. K. Akiyama, J.-C. Algaba, T. An, et al.,  Galaxies  {\bf 10}, 113  (2022).

2. J. Alves, C. Zucker, A.A. Goodman, et al., Nature {\bf 578}, 237 (2020).

3. R.A. Benjamin, E. Churchwell, B.L. Babler, et al., Astrophys. J. {\bf 630}, L149 (2005).

4. S.B. Bian, Y. Xu, J.J. Li, et al., Astron. J. {\bf 163}, 54 (2022).

5. V.V. Bobylev, A.T. Bajkova, MNRAS {\bf 408}, 1788 (2010).

6. V.V. Bobylev and  A.T. Bajkova, MNRAS {\bf 437}, 1549 (2014). 

7. V.V. Bobylev, O.I. Krisanova, A.T. Bajkova, Astron. Lett.  {\bf 46},  439, 2020.

8. V.V. Bobylev, A.T. Bajkova, Astron. Lett. {\bf 48}, 376 (2022).

9. V.V. Bobylev, A.T. Bajkova, and Yu.N. Mishurov, Astron. Lett. {\bf 48}, 434 (2022).

10. A. Cabrera-Lavers, C. Gonzalez-Fern\'andez, F. Garz\'on, P.L. Hammersley, and M. Lopez-Corredoira, Astron. Astrophys. {\bf 491}, 781 (2008).

11. J. Donada, F. Figueras, arXiv: 2111.04685 (2021).

12. G. Edenhofer, C. Zucker, P. Frank, et al., Astron. Astrophys. {\bf 685}, A82 (2024).

13. P. Englmaier, M. Pohl, and N. Bissantz,	arXiv: 0812.3491 (2008).

14. Gaia Collab. (T. Prusti, J.H.J. de Bruijne, A.G.A. Brown, et al.), Astron. Astrophys. {\bf 595}, 1 (2016).

15. Gaia Collab. (A. Vallenari, A.G.A. Brown, T. Prusti, et al.),  Astron. Astrophys. {\bf 674}, 1 (2023). 

16. P.A.B. Galli, L. Loinard, G.N. Ortiz-L\'eon,  et al., Astrophys. J. {\bf 859}, 33 (2018).

17. E. Griv, L.-G. Hou, I.-G. Jiang, and C.-C. Ngeow, MNRAS {\bf 464}, 4495 (2017).

18. E. Griv, M. Gedalin, and I.-G. Jiang, MNRAS {\bf 503}, 354 (2021).

19. M. Honma, T. Nagayama, K. Ando,  et al., PASJ {\bf 64}, 136 (2012).

20. L.J. Hyland , M.J. Reid, G. Orosz,  et al., Astrophys. J. {\bf 953}, 21 (2023).

21. L.J. Hyland,  S.P. Ellingsen, M.J. Reid, J. Kumar, and G. Orosz, {\it Cosmic Masers: Proper Motion toward the Next-Generation Large Projects}. Proceedings IAU Symposium No. 380, 2024, T. Hirota, H. Imai, K. Menten, and Y. Pihlstr\"om, eds.  (2024).

22. K. Immer, K.L.J. Rygl, Universe {\bf 8}, 390 (2022).

23. R. Konietzka, A.A. Goodman, C. Zucker, et al., Nature {\bf 628}, 62 (2024).

24. V. Krishnan,  S.P. Ellingsen,  M.J. Reid, et al., Astrophys. J. {\bf 805}, 129 (2015).

25. J.R.D. L\'epine, Yu.N.  Mishurov, and S.Yu. Dedikov, Astrophys. J. {\bf 546}, 234 (2021).

26. G.-X. Li, B.-Q. Chen, MNRAS {\bf 517}, L102 (2022).

27. Z. Li, J. Shen, O. Gerhard, and J.P. Clarke, Astrophys. J. {\bf 925}, 71 (2022).

28. X. Mai, B. Zhang, M.J. Reid,  et al., Astrophys. J. {\bf 949}, 10 (2023).

29. G.N. Ortiz-Le\'on,  L. Loinard,  M.A. Kounkel, et al., Astrophys. J. {\bf 834}, 141  (2017).

30. G.N. Ortiz-Le\'on, L. Loinard, S.A. Dzib,  et al.,   Astrophys. J. {\bf 865}, 73 (2018).

31. G.N. Ortiz-Le\'on, K.M. Menten, T. Kaminski, A. Brunthaler, M.J. Reid, and R. Tylenda,  Astron. Astrophys. {\bf 638}, 17 (2020).

32. G.N. Ortiz-Le\'on, S.A. Dzib, L. Loinard, Y. Gong, T. Pillai, and A. Plunkett,  Astrophys. Asyrophys. {\bf 673}, L1  (2023).

33. M. Pohl, P. Englmaier, and N. Bissantz, Asyrophys. J. {\bf 677}, 283 (2008).

34. A.S. Rastorguev,  N.D. Utkin, M.V. Zabolotskikh, et al., Astrophys. Bulletin {\bf 72}, 122 (2017).

35. M.J. Reid, K.M. Menten, X.W. Zheng, et al., Astrophys. J. {\bf 700}, 137 (2009).

36. M.J. Reid, N. Dame, K.M. Menten,  et al., Astrophys. J. {\bf 885}, 131 (2019).

37. N. Sakai, T. Nagayama, H. Nakanishi,  et al., Publ. Astron. Soc. Japan  {\bf 72}, 53 (2020).

38. N. Sakai, B. Zhang, S. Xu,  et al., Publ. Astron. Soc. Japan  {\bf 75}, 208 (2023).

39. L. Thulasidharan, E. D'Onghia, E. Poggio, et al.,   Astron. Astrophys. {\bf 660}, 12 (2022).

40. VERA collaboration, T. Hirota, T. Nagayama, M. Honma, et al.,  PASJ {\bf 70}, 51 (2020).

41. C. Wegg, O. Gerhard, and M. Portail, MNRAS {\bf 450}, 4050 (2015).

42. Y. Xu, S. B. Bian, M. J. Reid,  et al., Astrophys. J. Suppl. Ser. {\bf 253}, 9 (2021).

43. S.-Y. Yu, L.C. Ho, Astrophys. J. {\bf 871}, 194 (2019).

44. C. Zucker, J. Alves, A. Goodman, et al., Protostars and Planets VII, ASP Conf. Ser., Vol. 534, Proc. conf. held 10--15 April 2023 at Kyoto, Japan. Eds. S.-I. Inutsuka, Y. Aikawa, T. Muto, K. Tomida, and M. Tamura. San Francisco: Astron. Soc. Pacific, p.~43 (2023).
 }
\end{document}